\documentclass[12pt,nohyper,letterpaper]{JHEP3}         
\usepackage{amssymb,amsfonts}

\usepackage{epsfig}

\def\be{\begin{eqnarray}}
\def\ee{\end{eqnarray}}
\newcommand{\nn}{\nonumber}
\newcommand\para{\paragraph{}}
\newcommand{\ft}[2]{{\textstyle\frac{#1}{#2}}}
\newcommand{\eqn}[1]{(\ref{#1})}

\def\Dslash{\,\,{\raise.15ex\hbox{/}\mkern-12mu D}}
\def\Dbarslash{\,\,{\raise.15ex\hbox{/}\mkern-12mu {\bar D}}}
\def\delslash{\,\,{\raise.15ex\hbox{/}\mkern-9mu \partial}}
\def\delbarslash{\,\,{\raise.15ex\hbox{/}\mkern-9mu {\bar\partial}}}
\def\pslash{\,\,{\raise.15ex\hbox{/}\mkern-9mu p}}
\def\calDslash{\,\,{\raise.15ex\hbox{/}\mkern-12mu {\cal D}}}

\def\kslash{\,{\raise.15ex\hbox{/}\mkern-11mu k}}

\def\lae{\mathrel{\mathop{\smash{\lower .5 ex \hbox{$\stackrel<\sim$}}}}}
\def\lae{\mathrel{\mathop{\smash{\lower .5 ex \hbox{$\stackrel>\sim$}}}}}

\title{Flowing Between  Fermionic Fixed Points}
\author{Jo\~ao N. Laia and David Tong
 \\
Department of Applied Mathematics and Theoretical Physics, \\
University of Cambridge, UK\\
{\tt j.laia, d.tong@damtp.cam.ac.uk}
}

\abstract{We study holographic Wilsonian renormalization group flows for bulk spinor fields in AdS. We use this to compute beta functions for a number of double trace fermionic operators in the dual conformal field theory.}

\begin{document}
\pagestyle{plain} \setcounter{page}{1}
\newcounter{bean}
\baselineskip16pt

\section{Introduction}

In \cite{rg1,rg2}, a holographic framework to describe Wilsonian renormalization group (RG) flows was developed. While this framework preserves the Hamilton-Jacobi spirit of earlier approaches to holographic RG  \cite{deboer}, it differs by focussing attention on the boundary conditions experienced by   bulk fields.  Indeed, at heart, the Wilsonian RG flow equation can be viewed as an expression for how these boundary conditions evolve as the UV cut-off surface is varied. Aspects of this formalism were anticipated in \cite{iqliu,nickel} and further discussed in \cite{shin,nick}. 

\para
The story of boundary conditions for fields in anti-de Sitter space is intimately linked to the story of multi-trace operators in the dual conformal field theory. 
It has long been known that scalar fields in AdS whose mass lies in a certain window admit two different boundary conditions \cite{bf}. The implication for the AdS/CFT correspondence is that the same bulk dynamics can describe two different dual CFTs \cite{kw}.

\para
 These two CFTs are not unrelated. They share the same operator content, albeit with different operator dimensions, and their central charges\footnote{Here we adopt the language of $d=3+1$ CFTs for convenience. In other contexts $N^2 \rightarrow N^{3/2}$, $N^3$, $N^\#$\ldots}, which are of order large $N^2$, differ only at order one \cite{gubmit,gubkleb,hartrast}. Most pertinently, there is an RG flow between these two theories.  One of the theories has a  scalar operator ${\cal O}$ with dimension $\Delta < d/2$ where $d$ is the spacetime dimension of the CFT. This means that not only is ${\cal O}$ relevant, but ${\cal O}^2$ is also relevant. In the language of the 't Hooft large N  expansion, ${\cal O}$ is a single trace operator and ${\cal O}^2$ is a double trace operator.


\para
Turning on the single trace operator ${\cal O}$ will result in an RG flow which leads to a large backreaction on the bulk geometry and an order $N^2$ change in the central charge. This is not the flow of interest here. Instead, one finds a more subtle RG flow induced by turning on 
the operator ${\cal O}^2$. This can be shown to  affect only the boundary conditions of the bulk field and takes us from one  CFT to the other without affecting the bulk dynamics  at leading order in $N$ \cite{witten,micha}. One of the nice, concrete results of \cite{rg1,rg2} is the holographic computation of the beta function for the double trace operator which matches  earlier analysis based on large $N$ conformal perturbation theory \cite{rastelli} (see also \cite{roiban,kiritsis,vecchi}). However, this RG flow is a perilous descent, running down a mountain ridge with a  precipitous large $N$ drop to either side that can be avoided only by fine tuning the single trace ${\cal O}$ or, more naturally, by invoking symmetry if ${\cal O}$ is complex.

\para
The purpose of this paper is to implement the  techniques of Wilsonian holographic RG flow for bulk spinors, corresponding to fermionic operators $\Psi_\alpha$ in the dual conformal field theory. In a microscopic language, these are single trace composite operators typically of the form $\Psi_\alpha\sim {\rm Tr}(\phi\lambda_\alpha)$. Because of its Grassmann nature, one cannot perturb by the single trace $\Psi$ and the double trace perturbation  $\bar{\Psi}\Psi$ is the leading relevant deformation. Indeed, it is a simple matter to construct bottom-up bulk theories for which $\bar{\Psi}\Psi$ is the only relevant operator in the game. (Admittedly, from a top-down perspective, this may not be so natural and would require  
the anomalous dimensions of operators such as ${\rm Tr}\phi^2$ and ${\rm Tr}\,\bar{\lambda}\lambda$ to be large).

\para
Although conceptually similar to the scalar operators discussed in \cite{rg1,rg2}, the holographic RG computations for spinors have a slightly different flavour and this is reflected in the beta function for the double trace coupling which has a different structure from its scalar counterpart. This beta-function is given in equation \eqn{gbeta}.

\para
In Section 2, we review the Wilsonian holographic RG framework of  \cite{rg1,rg2}, adapted to fermions. 
Section 3 and 4 contain a fairly exhaustive discussion of fermionic double trace operators in AdS${}_4$ and AdS${}_5$ respectively. The first case we consider in Section 3.1 is a  $\bar{\Psi}\Psi$ deformation in $d=2+1$ dimensional CFT. We compute the double trace beta-functions both for vacuum RG flows and for flows in the presence of sources where higher derivative operators become important. In later sections we discuss various Majorana deformations and non-relativistic deformations. Despite the rather different gamma matrix structures involved, we find that the beta functions for all these cases remain essentially unchanged.

\section{Holographic Wilsonian RG for Fermions}

In this section we review  the holographic Wilsonian RG prescription of \cite{rg1,rg2}, with emphasis on bulk spinor fields. We work in a fixed background metric which, as $r\rightarrow 0$, asymptotes to AdS${}_{d+1}$,
\be ds^2 = \frac{L^2}{r^2}\left(dr^2 + \eta_{\mu\nu}dx^\mu dx^\nu\right)\label{ads}\ee
The $r\rightarrow 0$ limit of the geometry corresponds to the UV of the dual field theory. We will evaluate bulk dynamics with a UV cut-off in place at $r=\epsilon$. The corresponding action contains a bulk piece, together with a boundary term, 
\be S =\int_{r\geq \epsilon} d^{d+1}x \sqrt{-g}{\cal L}[\bar{\psi},\psi]+S_B[\bar{\psi},\psi,\epsilon]\label{action}\ee
The bulk Lagrangian describes a free Dirac spinor $\psi$,
\be
{\cal L}[\bar{\psi},\psi] = i\bar{\psi} \,\left[\frac{1}{2}\left(\Gamma^M\!\stackrel{\rightarrow}D_M-
\stackrel{\leftarrow}D_M\!\Gamma^M\right)-m\right]\psi\label{lag}\ee
Here the covariant derivative is $\stackrel{\rightarrow}D_M = 
\stackrel{\rightarrow}\partial_M + \ft14 \omega_{ab,M}\Gamma^{ab}$ where $\omega_{ab,M}$ is the bulk spin connection and $\Gamma^{ab} = \ft12 [\Gamma^a,\Gamma^b]$. In all these expressions, the capital index $M$ denotes bulk spacetime while $a,b$ denote bulk tangent space. In this paper we will restrict attention to AdS$_4$ and AdS$_5$ spacetimes; for both of these the irreducible representation of the gamma matrices $\Gamma^a$ is four-dimensional.
 
\para
The quantity $S_B[\bar{\psi},\psi;\epsilon]$ in \eqn{action}  is a boundary action, evaluated at $r=\epsilon$. It will determine the boundary conditions of the spinor and is the object of primary interest in this paper. The necessity of this term is easily seen by varying the action. The Dirac equation of motion,
\be \left(\Gamma^MD_M-m\right)\psi=0\label{dirac}\ee
only arises as an extremum of the action if accompanied by a successful cancelation of the boundary terms.  For most studies of spinors in AdS, this is achieved by imposing Dirichlet conditions for half of the spinor fields. Here, following \cite{rg1,rg2}, we instead choose to work with Neumann boundary conditions,
\be \Pi = \frac{\delta S_B}{\delta\bar{\psi}}\ \ \ ,\ \ \ \bar{\,\Pi} = \frac{\delta S_B}{\delta \psi}\label{neumann}\ee
where $\Pi$, the canonical momenta of the spinor field in the radial direction, is defined as
\be \Pi \equiv -\frac{i}{2}\sqrt{-g}\,e^r_{a}\,\Gamma^a\psi\ \ \ ,\ \ \ \bar{\Pi} = -\frac{i}{2}\sqrt{-g}\bar{\psi}\,e^r_a\,\Gamma^a\label{pi}\ee
Here we have endowed the gamma matrices with tangent space indices at the expense of introducing the vielbein, $e^{r}_a$.

\para
In pure AdS,  with metric \eqn{ads}, the vielbein is  simply $e^r_a = \delta_{a4}\, r/L$ where $a=4$ is understood as the tangent space index that points in the radial AdS direction. The boundary term which preserves conformal invariance is,
\be S_B = \frac{i}{2}\int_{r=\epsilon}\sqrt{-\gamma}\,f\bar{\psi}\psi\label{f1}\ee
where $f$ is a constant which we shall determine shortly and $\gamma$ is the determinant of the induced boundary metric, $\gamma=gg^{rr}$. The Neumann boundary condition \eqn{neumann} then reduces to
\be {\cal M}\psi = 0\ \ \ \ {\rm with}\ \ \ \ {\cal M} = f{\bf 1} + \Gamma^4\label{confbc}\ee
The fact that this is a conformal boundary condition is reflected in the lack of $\epsilon$ dependence. However, we must still determine the coefficient $f$. There are only two consistent possibilities. The reason for this can be traced back to the fact that bulk spinors obey first order equations of motion. This means that the different components of $\psi$ on the boundary $r=\epsilon$  correspond to both ``position" and ``momentum" variables, as can be seen in the explicit expressions for the canonical momentum \eqn{pi}. But we don't want to fix both position and momentum on the boundary. That would be overkill. We only want to fix half of the components of $\psi$ which means that the $4\times 4$ matrix ${\cal M}$ obeys
\be {\rm rank}\,{\cal M}=2\label{mconf}\ee
Restrictions of this type will play an important role shortly when we come to understand how the boundary conditions evolve under RG flows. For now, we wish to solve this condition for the simple matrix ${\cal M}$ given in \eqn{confbc}. Using the fact that $(\Gamma^4)^2=\bf{1}$, ${\cal M}$ has rank 2 if and only if  $f=\pm 1$. The choice $f=+1$ sets $\psi_+ \equiv \ft12({\bf 1}+\Gamma^4)\psi=0$ on the boundary while the choice $f=-1$ sets $\psi_-=\ft12({\bf 1}-\Gamma^4)\psi=0$

\para
Which of these two conditions, $\psi_+=0$ or $\psi_-=0$, is acceptable depends on the mass $m$ of the bulk spinor. The question is one of normalizability: if we choose to fix $\psi_+=0$, then we had better make sure that $\psi_-$ is actually able to move. To see whether this is the case, we must look a little closer at the solutions of the Dirac equation \eqn{dirac} near the boundary. These too can be written conveniently in terms of eigenspinors of $\Gamma^4$. Working in Fourier modes for the boundary directions, $\psi(x,r) = \psi(k;r)e^{ik\cdot x}$, the solution close to the boundary, $r\rightarrow 0$, takes the form
\be \psi_- (k;r)&=& A(k) r^{d/2-mL}+B(k)r^{d/2+mL+1} +\ldots\nn\\
\psi_+ (k;r)&=& D(k)r^{d/2+mL} +C(k)r^{d/2-mL+1}+  \ldots \label{asym}\ee
The spinors $A$, $B$, $C$ and $D$ are related by the conditions
\be D = -\frac{i\Gamma\cdot k}{k^2}(2m+1)B\ \ \ ,\ \ \ C=\frac{i\Gamma\cdot k}{2m-1}A\nn\ee
Although the bulk action \eqn{action} has only boundary contributions when evaluated on a solution to the Dirac equation \eqn{dirac}, the question at hand here is whether the energy for a given mode is finite. Evaluated upon the asymptotic solutions \eqn{asym}, this energy takes the form
\be {\cal E} \sim \int_{r\geq \epsilon}dr\ \frac{1}{r^{d+1}}\left[\bar{A}C\,r^{d-2mL+1} - \bar{D}B\,r^{d+2mL+1}\right
] \nn\ee
>From this we can read off the result. For $mL  \geq 1/2$, the $\bar{D}B$ term is  normalizable but the $\bar{A}C$ term is non-normalizable. This means that $A$ cannot fluctuate dynamically without infinite cost of energy and we are obliged to set $\psi_-=0$ on the boundary. For $mL \leq -1/2$, this story is reversed and we must set $\psi_+=0$ on the boundary. But for $-1/2 < mL < 1/2$, both terms are normalizable \cite{iliu}. Here we have a choice.

\para
Our interest in this paper is in the window $|mL| < 1/2$. We will restrict attention to $mL\geq0$ since the case $mL<0$ is trivially related by exchanging $\psi_+$ and $\psi_-$. The two different boundary conditions for the spinor $\psi$ result in two different boundary conformal field theories. They are usually referred to as the standard and alternative quantization and we now review their properties.
\\{\ }\\
\noindent
\underline{Standard Quantization}: $f=-1$. The condition \eqn{confbc} sets $\psi_-=0$ on the boundary, fixing the most divergent term $A$ to zero. More generally, the addition of  linear piece to the boundary term
\be S_B = -\frac{i}{2}\int_{r=\epsilon}\sqrt{-\gamma} \ \left[\bar{\psi}\psi + \bar{\eta}\psi + \bar{\psi}\eta\right]\nn\ee
sets $A=\epsilon^{mL-d/2}\eta$ and $\eta$ is  interpretated as a source  for the boundary fermionic operator $\Psi$. Correspondingly, the spinor $D$ in \eqn{asym} has the interpretation as the expectation value in the presence of this source:  $D\sim \langle\Psi\rangle_A$. The dimension of $\Psi$, which can be read from the power-law fall-offs of these coefficients, is
\be \Delta_+=\frac{d}{2}+mL\nn\ee
\\ 
\noindent
\underline{Alternative Quantization}: $f=1$. The condition \eqn{confbc} now sets $\psi_+=0$. The roles of $\psi_-$ and $\psi_+$ are now flipped, with the bulk spinor $D$ interpreted as the source and $A\sim\langle\Psi\rangle_D$ as the response. Correspondingly, the boundary fermionic operator $\Psi$ has dimension
\be \Delta_-=\frac{d}{2}-mL\nn\ee
Importantly, in the alternative quantization, a double trace operator constructed from bilinears of $\Psi$ has dimension $d-2mL$ and, for $mL>0$, is relevant. The goal of this paper is to understand the RG flows induced upon turning on such relevant operators. These RG flows, some aspects of which were previously examined in \cite{allais}, typically take us from the alternative quantization to the standard quantization. For this reason, the alternative choice ($f=+1$) is often referred to as the UV CFT while the standard choice ($f=-1$) is referred to as the IR CFT.

\subsubsection*{RG Flow Equations}

We now turn to the Wilsonian holographic RG framework introduced in \cite{rg1,rg2}.  When performing calculations in AdS/CFT, the bulk equations of motion are solved subject to two boundary conditions: one in the UV, at some cut-off $r=\epsilon$, and the other in the IR. This solution is then fed back into the bulk action, from which correlations function of the boundary theory can be evaluated.

\para
 The strategy of \cite{rg1,rg2} is to examine how this recipe changes as 
we vary $\epsilon$, corresponding to integrating out high-energy modes in 
the boundary theory. We insist that (suitable) physical quantities, which 
are encoded in the on-shell bulk action, are independent of the choice of 
$\epsilon$. This can be achieved only at the cost of changing the boundary 
conditions imposed on fields at $r=\epsilon$ which, in turn, requires that the boundary contribution $S_B$ to the bulk action varies with $\epsilon$. The Wilsonian RG flow equation is an expression dictating the change of $S_B$. 

\para
It is a simple matter to derive the RG flow equation for fermions with action \eqn{action}. We have
\be S[\epsilon+\delta\epsilon]-S[\epsilon]  &=&  \int_{r=\epsilon+\delta\epsilon}^{r=\epsilon}d^{d+1}x\, \sqrt{-g}\,{\cal L}[\psi]
+S_B[\psi(x,\epsilon+\delta\epsilon);\epsilon+\delta\epsilon] - S_B[\psi(\epsilon);\epsilon] \nn\ee
Expanding to leading order in $\delta \epsilon$ (and recalling the Grassmannian nature of $\psi$ to get the minus signs right), gives
\be
\frac{dS}{d\epsilon} =  - \int_{r=\epsilon}d^dx\ \sqrt{-g}\,{\cal L}[\psi] +\int_{r=\epsilon}d^dx\left(-\frac{\delta S_B}{\delta\psi}\partial_r\psi +\partial_r\bar{\psi}\,\frac{\delta S_B}{\delta \bar{\psi}}\right)
 +\left.\frac{\partial S_B}{\partial \epsilon}\right|_{r=\epsilon} \nn\ee
At this point something rather nice happens. All terms above are evaluated at $r=\epsilon$ where the Neumann boundary conditions \eqn{neumann}  allow us to replace  $\delta S_B/\delta\psi = \bar{\Pi}$. The variation of the on-shell bulk action then reduces to the simple expression
\be \frac{dS}{d\epsilon} = \int_{r=\epsilon} d^dx \ {\cal H} + \left.\frac{\partial S_B}{\partial \epsilon}\right|_{r=\epsilon} \nn\ee
where ${\cal H}$ is the Hamiltonian for {\it radial} evolution. Because the action is first order, the explicit $\bar{\Pi}\partial_r\psi$ terms cancel with those in ${\cal L}$ in the Hamiltonian,
\be {\cal H} &=& -\bar{\Pi}\,\partial_r\psi + \partial_r\bar{\psi}\,{\Pi} -\sqrt{-g}{\cal L}\nn\\ &=& -\frac{i}{2}\sqrt{-g}\left[e^\mu_a\,\bar{\psi}\,(\Gamma^a\!\stackrel{\rightarrow}D_\mu-\stackrel{\leftarrow}D_\mu\!\Gamma^a)\psi-2m\bar{\psi}\psi\right]\nn\\ &=&  -\frac{i}{2}\sqrt{-g}\left[e^\mu_a\,\bar{\psi}\left(2\Gamma^a\partial_\mu + \frac{1}{4}\omega_{bc,\mu}\{\Gamma^a,\Gamma^{bc}\}\right)\psi-2m\bar{\psi}\psi\right]\nn \ee
The vielbein in this expression reflects the fact that all gamma matrices carry tangent space indices, now running only over boundary directions. In pure AdS, this expression simplifies further as the spin connection term vanishes and the vielbein is $e^\mu_a =  (r/L)\delta^\mu_a$. 

\para
As mentioned above, because the Dirac action is first order the Hamiltonian does not depend on the radial momentum $\Pi$.  This is in contrast to the bosonic fields studied in \cite{rg1,rg2}, where the quadratic terms ${\cal H}_{\rm bosonic}\sim\Pi^2$ gave rise to most of the interesting physics of the RG flow. For example, the structure of the beta-function for scalar operators can be traced to these terms. We shall see in the next section that the non-trivial beta-functions for fermionic operators arise in a slightly different manner.

\para
Finally, physics of the boundary theory is guaranteed to be independent of the choice of UV cut-off if $dS/d\epsilon=0$. This is to be understood as a flow equation for the boundary action, 
\be \frac{\partial S_B}{\partial \epsilon} = -\int_{r=\epsilon} d^dx\ {\cal H}\label{rgflow}\ee
The rest of this paper is devoted to a study of this equation for various fermionic RG flows in AdS${}_4$ and AdS${}_5$.

\section{Spinors in AdS${}_4$}

A bulk Dirac spinor $\psi$ in AdS${}_4$ is dual to a Dirac fermion operator in the $d=2+1$ dimensional boundary. We denote this two component, complex spinor operator as $\Psi_\alpha$, $\alpha=1,2$. We will work with the gamma matrices, 
\be \Gamma^\mu = \left(\begin{array}{cc} 0 & \gamma^\mu \\ \gamma^\mu & 0 \end{array}\right),\  \mu =0,1,2\ \ \ ,\ \ \ \
\Gamma^4 = \left(\begin{array}{cc} 1 & 0 \\ 0 & -1 \end{array}\right) \ \ \ ,\ \ \ \Gamma^5 = \left(\begin{array}{cc} 0 & 1 \\ -1 & 0 \end{array}\right)\nn\ee
where $\gamma^\mu=(i\sigma^3,\sigma^1,\sigma^2)$ furnish a representation of the 3d Clifford algebra. 

\para
We start in the UV CFT with $f=+1$. Our game plan is to follow the RG flow that is induced by  relevant double trace operator  formed from a fermion bilinear. There are two such Lorentz invariant bilinears that can be constructed from a Dirac spinor $\Psi$ and there is also a  Lorentz violating term of interest. Each of these has (at large $N$) dimension $2\Delta_-=3-2mL$ and is relevant for $0<mL<1/2$. (The upper bound coincides with the unitarity bound). The deformations we will consider are:
%

%
\begin{itemize}
\item Dirac Deformation $\bar{\Psi}\Psi$: If $\Psi$ were a free fermion, rather than a composite operator, this would be a Dirac mass term. This preserves both Lorentz symmetry and charge conjugation, but breaks parity.
\item Majorana Deformation $\bar{\Psi}^{(c)}\Psi + {\rm h.c.}$: Here the conjugate spinor is defined to be $\Psi^{(c)} = C\Psi^\star$ where, in the conventions above, the 3d charge conjugation matrix is  $C=\sigma^1$. 
For a free fermion, this deformation is a Majorana mass. It preserves Lorentz symmetry and parity, but at the cost of  breaking the $U(1)$ global symmetry which rotates the phase of $\Psi$. 
\item ``Chemical Potential" Deformation  $\bar{\Psi}\gamma^0\Psi$: This breaks Lorentz invariance. It preserves parity, but breaks charge conjugation symmetry. We stress that because $\Psi$ is a composite operator --- typically something like $\Psi_\alpha = {\rm Tr}(\phi\lambda_\alpha)$ --- this deformation is {\it not} literally a chemical potential of the boundary theory. (As always, that is turned on by providing a source for the temporal component of a current  operator). 
\end{itemize}

\subsection{Dirac Deformation}

We start by considering a vacuum flow from the UV CFT. We set sources to zero and deform the theory by the relevant operator 
\be \Delta S_{\rm Dirac} = i\int \frac{d^3k}{(2\pi)^3}\ 
\zeta\bar{\Psi}(k){\Psi}(k)\label{minusimp}\ee
with constant $\zeta$.   We propose that this deformation is modelled by 
adding to the boundary term \eqn{f1} the expression $g\bar{\psi}\Gamma^5\psi$, 
where the relationship between $g$ and $\zeta$ will be derived shortly.  
The full boundary term is then given by
\be S_B=\frac{i}{2}\int_{r=\epsilon}\sqrt{-\gamma}\left[f(\epsilon) \bar{\psi}\psi + g(\epsilon)\bar{\psi}\Gamma^5\psi\right]\label{g1}\ee
There are a number of consistency checks for this proposal. First, note that the symmetries agree: the term $\psi\Gamma^5\psi$ breaks parity in the bulk, mirroring $\bar{\Psi}\Psi$ in the boundary theory. Secondly, the proposal morally follows the prescription of \cite{witten,micha} for scalar operators. To see this recall that, in the UV CFT, the spinor $A$ in the expansion \eqn{asym} is related to the boundary operator: $A\sim\langle\Psi\rangle$. The term $\bar{\psi}\Gamma^5\psi\sim A^\dagger\gamma^0A$ then contains two copies of the operator, which is the key property of scalar double trace perturbations given in  \cite{witten,micha}. In contrast, the term $\bar{\psi}\psi \sim A^\dagger \gamma^0D$ does not have the structure appropriate for a double trace perturbation. 

\para
We must now examine afresh the requirement that $S_B$ provides a good boundary condition. The Neumann boundary condition \eqn{neumann} becomes ${\cal M}\psi=0$ with 
\be {\cal M} = f(\epsilon){\bf 1} + \Gamma^4 + g(\epsilon)\Gamma^5\label{m2}\ee
As discussed previously, we require that only half the spinor components are fixed on the boundary, so that ${\rm rank}\,{\cal M}=2$. This only holds if $f$ and $g$ obey
\be f(\epsilon)^2 + g(\epsilon)^2 =1\label{fg}\ee
We now insert \eqn{g1} in the RG flow equation \eqn{rgflow}, specialising to pure AdS. As mentioned previously, the spin connection term vanishes in AdS. For our vacuum flow, all bulk fields are constant in the boundary directions, $\partial_\mu \psi=0$, ensuring that the gradient terms also vanish.  We're left with,
\be \frac{\partial}{\partial\epsilon}\left(\frac{L^d}{\epsilon^d}f\right)\bar{\psi}\psi + \frac{\partial}{\partial\epsilon} \left(\frac{L^d}{\epsilon^d}g\right)\bar{\psi}\Gamma^5\psi = -2\frac{L^{d+1}}{\epsilon^{d+1}}m \bar{\psi}\psi\label{lewis}\ee
At this point, it pays to employ the boundary condition ${\cal M}\psi=0$, with ${\cal M}$ given in \eqn{m2}. It is a simple matter to show that this implies $\bar{\psi}\Gamma^4\psi=0$ and $f\bar{\psi}\psi + g\bar{\psi}\Gamma^5\psi=0$. Using both of these, together with the constraint \eqn{fg}, the flow equation \eqn{lewis} can be massaged into a beta function for $f(\epsilon)$,
\be \epsilon\frac{\partial f}{\partial\epsilon} = -2mL (1-f^2)\nn\ee
with solution
\be f(\epsilon) = \frac{4-\zeta^2\epsilon^{4mL}}{4+\zeta^2\epsilon^{4mL}}\label{betaf}\ee
Here the coefficient $\zeta$ of \eqn{minusimp} has made a re-appearance as an 
integration constant. This solution behaves as expected, interpolating from the UV CFT with  $f=+1$ at $\epsilon =0$ to the IR CFT with $f=-1$ as 
$\epsilon \rightarrow \infty$. 

\para
However, the direct meaning of $f$ in the boundary CFT is unclear. Instead, it 
is $g(\epsilon)$ which should be thought of as the coefficient  of the double 
trace operator. For $f>0$, its beta function is given by
\be \epsilon\frac{\partial g}{\partial \epsilon} = 
2mLg\sqrt{1-g^2}\label{gbeta}\ee
The coefficient $g$ increases monotonically until $g=1$ and $f=0$, at which point 
$f$ changes sign and the beta function is replaced by
\be \epsilon\frac{\partial g}{\partial \epsilon} = 
-2mL g\sqrt{1-g^2}\nn\ee
The solution for the full flow is given by
\be g(\epsilon) = 
\frac{4\zeta\epsilon^{2mL}}{4+\zeta^2\epsilon^{4mL}}\nn\ee
At the start of the RG flow, $g$ is related to the coefficient of the double trace operator by $g\approx \zeta \epsilon^{2mL}$. Since $g$ is dimensionless, the integration constant $\zeta$ must have dimension $2mL$ as befits the deformation \eqn{minusimp}. 
In contrast to $f$, the flow of the coupling $g$ is not monotonic. Instead, $g\rightarrow 0$ in both UV and IR. Indeed, this is necessary to ensure  that the fixed points are parity invariant although the flow itself is not.

\para
It is worth comparing the beta function for the fermionic coupling with that of the scalar double trace operator computed in \cite{rg1,rg2}. In the scalar case, the beta function was only quadratic in coupling, similar to our beta function \eqn{betaf}. This implies that perturbation  terminates at one-loop, a fact that had been previously derived from large $N$-ology \cite{rastelli}. Note that ``one-loop" here refers to perturbation theory in the double trace coefficient; not perturbation theory in the 't Hooft coupling. 
In contrast, the beta function \eqn{gbeta} suggests that the double trace perturbation \eqn{minusimp} receives divergent $2n$ loop contributions for all $n$. The lack of odd powers of $\zeta$ can be seen from gamma-matrix algebra alone. It would be interesting to understand how this result tallies with large $N$ counting in conformal perturbation theory.

\subsection*{Higher Derivative Operators}

To understand the flow of more general excitations, we must nudge the theory 
away from the ground state. As usual, this is achieved by introducing sources  
in the boundary Lagrangian,
\be  \Delta S_{\rm Source} =  i\int \frac{d^3k}{(2\pi)^3}\ \,
 \bar{\eta}(k)\Psi(k) + \bar{\Psi}(k)\eta(k)  \label{source}\ee
One of the lessons of \cite{rg1,rg2} is that while the sources break translational invariance, the multi-trace operators that they induce do not. Rather, the translational symmetry breaking manifests itself in the need to turn on descendant operators containing higher derivative terms. Anticipating this possibility, we should consider all possible higher derivative operators, $\bar{\Psi}\!\delslash\Psi$, $\bar{\Psi}\partial^2\Psi$, $\bar{\Psi}\partial^2\!\delslash\Psi$ and so.

%
%

\para
In the boundary action $S_B$, operators of the form $\bar{\Psi}\partial^{2n}\Psi$ can be conveniently packaged by promoting the constant $g$ to a function of $k^2\equiv k_\mu k^\mu = -\omega^2 + \vec{k}^2$. Successive terms in the Taylor expansion of $g(k^2)$ are related to the coefficients of the tower of higher derivative operators.  However, to take into account operators of the form 
$\bar{\Psi}\partial^{2n}\!\delslash\Psi$, we need to introduce a different boundary term whose coefficient is again a function $h(k^2)$. All told, the general form of the boundary term is
\be S_B = \frac{i}{2}\int_{r=\epsilon} \sqrt{-\gamma}\left[f(k^2)\,\bar{\psi}\psi + g(k^2)\,\bar{\psi}\Gamma^5\psi + i h(k^2)\,\bar{\psi}\kslash \psi -\bar{\eta}\psi - \bar{\psi}\eta\right]\label{mets}\ee
Here  $\kslash \equiv k_\mu \Gamma^\mu$ sums over the bulk gamma matrices tangent to the boundary. Notice that the spinor structure of the $h(k^2)$ term is of the same form as the $g(k^2)$ term. It includes $A^\dagger k_\mu\gamma^\mu A$, rather than the $A^\dagger\cdot D$ structure of the $f(k^2)$ term. The Neumann boundary condition on the spinor becomes
\be {\cal M}\psi=\eta
\label{meta}\ee
with
\be {\cal M} = f {\bf 1}+\Gamma^4+g\Gamma^5 + ih\,\kslash\label{m4tunnel}\ee
There is a subtlety here. The source  in \eqn{source} is a 2-component spinor while $\eta$ in \eqn{mets} is a four-component spinor. This discrepancy is resolved by the requirement \eqn{meta} that $\eta$ lies in the image of the rank 2 matrix ${\cal M}$. The source must be embedded in the appropriate two-dimensional subspace. This embedding rotates under the RG flow.

\para
Again, the requirement that the boundary condition \eqn{meta} doesn't over-fix  means that we must have ${\rm rank}\,{\cal M}=2$. This holds providing
\be f^2+g^2 + h^2k^2 =1\label{f15}\ee
Following the previous steps, we evaluate the RG flow equation \eqn{rgflow} evaluated on \eqn{mets}, this time retaining the kinetic term in ${\cal H}$. We make liberal use of the condition \eqn{meta} to arrive at
\be \epsilon\frac{\partial f}{\partial\epsilon}\bar{\psi}\psi + \epsilon\frac{\partial g}{\partial\epsilon}\bar{\psi}\Gamma^5\psi + i \epsilon\frac{\partial h}{\partial\epsilon}\bar{\psi}\kslash\psi -\epsilon\frac{\partial \bar{\eta}}{\partial\epsilon}\psi - \epsilon\bar{\psi}\frac{\partial \eta}{\partial\epsilon}\ \ \ \ \ \ \ \  \nn\\ = 2i\epsilon\bar{\psi} \kslash\psi - 2mL\bar{\psi}\psi- \frac{3}{2}(\bar{\eta}\psi + \bar{\psi}\eta)\nn\ee
This equation contains four distinct algebraic bilinears --- $\bar{\psi}\psi$, $\bar{\psi}\Gamma^5\psi$, $\bar{\psi}\kslash\psi$ and $\bar{\eta}\psi$ --- but there exists a single relationship between them which follows from \eqn{meta}: 
\be
f\bar{\psi}\psi + g\bar{\psi}\Gamma^5\psi + ih\bar{\psi}\kslash\psi = \frac{1}{2}(\bar{\eta}\psi+\bar{\psi}\eta)\nn\ee
We use this to eliminate one bilinear. The coefficients of the remaining three then provide us with the beta-functions. In this manner, we derive the beta-function for the source
\be \epsilon\frac{\partial \eta}{\partial\epsilon} = \left(\frac{3}{2}+\frac{mL}{f}+\frac{\epsilon}{2f}\frac{\partial f}{\partial\epsilon}\right)\eta\label{sourcebeta}\ee
Notice that for $f=\pm 1$, this result correctly reproduces the dimension: $\Delta[\eta] = 3-\Delta_{\mp}$. Away from the fixed points, this coefficient has a natural interpretation as ($3$ minus) the anomalous dimension of the operator $\Psi$. 

\para
The beta functions for the two double trace couplings $g$ and $h$ can be written implicitly as 
\be f\epsilon\frac{\partial g}{\partial\epsilon}-g\epsilon\frac{\partial f}{\partial\epsilon} &=&2mLg \label{betamax}\\ 
f\epsilon\frac{\partial h}{\partial\epsilon} -h\epsilon\frac{\partial f}{\partial\epsilon} &=& 2\epsilon f+2mLh\label{vhs}\ee
These equations should be understood subject to the constraint \eqn{f15} which can be used to eliminate $f(k^2)$. The second of these beta functions \eqn{vhs} arises from the coefficient of the  $\bar{\psi}\kslash\psi$ terms and, in the absence of any sources, can be ignored because $k=0$. In that case, the first beta function \eqn{betamax} reduces to \eqn{gbeta} that we saw earlier.

\subsection{Majorana Deformation}\label{majsec}

We turn now to the Majorana deformation in the boundary theory. 
We firstly return to the vacuum flow, with no explicit sources and only 
a double trace deformation
\be \Delta S_{Majorana} = i\int \frac{d^3k}{(2\pi)^3}\ \,
 \left[\zeta\bar{\Psi}^{(c)}(-k) \Psi(k) + \zeta^\star 
\bar{\Psi}(-k)\Psi^{(c)}(k)\right]\label{office}\ee
where we remind the reader that our conventions are $\Psi^{(c)} = C\Psi$ with $C=\sigma^1$. Notice that, in contrast to \eqn{minusimp}, the arguments of the operators are $k$ and $-k$. This extra minus sign follows simply from the definition of the  Fourier transform but is essential to get right in what follows.  In general, the coefficient  $\zeta\in {\bf C}$ but its phase can be eliminated by a rotation of the phase of $\Psi$ and we choose to invoke this freedom to set $\zeta\in {\bf R}$. 

\para
The charge conjugation of the bulk Dirac spinor is given by $\psi^{(c)} = {\cal C}\psi^\star$ with ${\cal C}=\Gamma^0\Gamma^2$. We propose that the Majorana deformation is implemented by  the boundary term
\be S_B=  \frac{i}{2}\int_{r=\epsilon} \sqrt{-\gamma}\left[f\bar{\psi}\psi + \frac{1}{2}g\bar{\psi}^{(c)}\Gamma^5\psi +\frac{1}{2}g\bar{\psi}\Gamma^5\psi^{(c)}\right]\label{deadonship}\ee
The arguments for this boundary term are analogous to those of the Dirac deformation: it is the unique deformation preserving (and breaking) the same symmetries as \eqn{office}. Moreover, the leading order behaviour near the boundary takes the expected  form $\bar{\psi}^{(c)}\Gamma^5\psi \sim D^T\sigma^1 D$. Note that, as is usually the case for Majorana masses, we must now be careful to treat $\psi$ as a Grassmann valued object. (This is less important in the case of Dirac deformations since the calculations are insensitive to this sign). 

\para
The Neumann boundary condition for \eqn{deadonship}  reads
\be (f{\bf 1} + \Gamma^4)\psi + g\Gamma^5\psi^{(c)} = 0\label{overboard}\ee
We again wish to find the conditions under which \eqn{overboard} fixes only half of the spinor degrees of freedom on the boundary. There are a number of ways to go about this. Perhaps the most straightforward is to translate \eqn{overboard} into an $8\times 8$ matrix equation acting on the components of ${\rm Re}\psi$ and ${\rm Im}\psi$ and require that four of the eigenvalues vanish. However, a computationally easier method (which will prove useful later) is to instead insist that the complex conjugate of \eqn{overboard} is actually the same equation. The complex conjugate is
\be (f{\bf 1} + \Gamma^4)\psi^{(c)} + g\Gamma^5\psi = 0\nn\ee
It is simple to put this in the form of \eqn{overboard}; we need only multiply by $(f+\Gamma^4)\Gamma^5$. But for the resulting equation to be identical to \eqn{overboard}, we still need to impose one further condition: it is the same restriction  \eqn{fg} that arose for the Dirac deformation: $f^2+g^2=1$. The remainder of the calculation proceeds similarly to the Dirac deformation and one finds the same beta-function, 
\be \epsilon\frac{\partial g}{\partial \epsilon} = 2mL g\sqrt{1-g^2}\label{splash}\ee
As with the Dirac deformation, this comes with a flip of a minus sign once $g$ reaches 1. 

\para
The equivalance of the beta functions for Dirac and Majorana deformations is somewhat surprising given 
that the flows themselves are very different. The Majorana flow, for example, breaks the $U(1)$ charge 
symmetry at all energy scales except at the end points. Nonetheless, the rate at which the RG flows take 
place is the same. This suggests that the beta function is a general property of fermionic double trace operators in large $N$ theories.

%
%

\subsubsection*{Higher Derivative Operators}

We now derive the  beta functions for the Majorana deformation in the presence of a source \eqn{source}. 
As in the Dirac deformation, we must anticipate the need to include higher derivative operators, $\bar{\Psi}\partial^{2n}\Psi$ and $\bar{\psi}\!\delslash\partial^{2n}\Psi$. With this in mind, the boundary term 
becomes
\be S_B=  \frac{i}{2}\int_{r=\epsilon} \sqrt{-\gamma}\left[f\bar{\psi}\psi + \frac{1}{2}g\left(\bar{\psi}^{(c)}\Gamma^5\psi + \bar{\psi}\Gamma^5\psi^{(c)}\right) + ih\bar{\psi}\kslash\psi -\bar{\eta}\psi-\bar{\psi}\eta\right]\ \ \ \ \ \ \ \label{fml}\ee
where $f$, $g$ and $h$ are again taken to be function of $k^2$. 

\para
The Neumann boundary condition imposed  at $r=\epsilon$ is,
\be(f{\bf 1}+\Gamma^4 + ih\kslash)\psi(k) +g\Gamma^5\psi^{(c)}(-k)=\eta(k)
\label{omfg}\ee
Here we have been careful to make the arguments of $\psi$ and $\psi^{(c)}$ explicit. The difference in minus sign can be traced to that in \eqn{office}. (It is also present, although implicit, in \eqn{fml}). 
As in the previous examples, we require that \eqn{omfg} fixes only half the spinor components and the simplest way to impose this is to again insist that  \eqn{omfg} and its complex conjugate are the same equation. The complex conjugate equation is
\be(f{\bf 1}+\Gamma^4 - ih\kslash)\psi^{(c)}(k) + g\Gamma^5\psi(-k)=\eta^{(c)}(k)
\nn\ee
Inverting the matrices, this can be written as
\be g (f{\bf 1}+\Gamma^4 + ih\kslash)\psi(k) - (f^2+h^2k^2-1)\Gamma^5\psi^{(c)}(-k) \ \ \ \ \ \ \ \ \ \ \ \ \ \  \nn\\ 
=-\Gamma^5(f{\bf 1}-\Gamma^4-ih \kslash)
\eta^{(c)}(-k)\nn\ee
This equation coincides with \eqn{omfg} is we impose the algebraic constraint
\be f^2 + g^2 + h^2k^2 = 1\nn\ee
which happily is the same requirement \eqn{f15} that we found for the Dirac deformation. We further have a condition on $\eta$,
\be g\Gamma^5\eta(k) = (f{\bf 1}-\Gamma^4 -ih\kslash)\eta^{(c)}(-k)\nn\ee
This again arises because $\eta$ in the expression above is a four component bulk spinor, while the source in \eqn{source} is a two component boundary object. The requirement above is entirely analogous to the statement we made after \eqn{meta} that $\eta$ should be in the image of ${\cal M}$.

\para
Evaluating the RG flow equation \eqn{rgflow} on the boundary term \eqn{fml} results in a long expression containing four different fermi bilinears --- $\bar{\psi}\psi$, $\bar{\psi}\kslash\psi$, $(\bar{\psi}\Gamma^5\psi^{(c)} + \bar{\psi}^{(c)}\Gamma^5\psi)$, and $\bar{\eta}\psi$. As with the Dirac deformation, one of these can be eliminated through the constraint \eqn{omfg} which  can be shown to imply,
\be
\bar{\psi}(k)(f + ih\kslash)\psi(k)  +\frac{g}{2}\left[\bar{\psi}(k)\Gamma^5\psi^{(c)}(-k) + \bar{\psi}^{(c)}(-k)\Gamma^5\psi(k)\right] = \frac{1}{2}\left[\bar{\eta}(k)\psi(k)+\bar{\psi}(k)\eta(k)\right]\nn\ee
The beta functions can then be read off from the remaining three bi-linears. They are exactly the same as we found for the Dirac equation: the beta function for the sources is given by \eqn{sourcebeta} while the beta-functions for $g$ and $h$ are \eqn{betamax} and \eqn{vhs} respectively.

\subsection{Chemical Potential Deformation}

In \cite{flatband}, we pointed out that spinors in AdS${}_4$ allow for two further scale invariant boundary conditions. These break Lorentz symmetry of the boundary theory but preserve rotational symmetry. In this section, we explore the RG flows to these fixed points, starting once again from the UV CFT. 

\para
As anticipated in \cite{flatband}, one can reach the non-relativistic fixed point by turning on a combination of the Dirac deformation $\bar{\Psi}\Psi$ and the ``chemical potential" $\bar{\Psi}\gamma^0\Psi$. We parameterize linear combinations of these two deformations thus,
\be \Delta S_{\rm Non-Rel} = i\int \frac{d^3k}{(2\pi)^3}\ \,\zeta_+\left(\bar{\Psi}\Psi + i\bar{\Psi}\gamma^0\Psi\right)+ \zeta_-\left(\bar{\Psi}\Psi - i\bar{\Psi}\gamma^0\Psi\right)
\label{goingsouth}\ee
>From the bulk perspective, it is useful to introduce the two projection operators 
\be P_\pm = \frac{1}{2}\left(1\pm i\Gamma^1\Gamma^2\right)\nn\ee
The double trace perturbation \eqn{goingsouth} is then implemented by the boundary term
\be S_B=\frac{i}{2}\int_{r=\epsilon}\sqrt{-\gamma}\left[f_-\bar{\psi}P_-\psi +f_+ \bar{\psi}P_+\psi + g_- \bar{\psi}\Gamma^5P_-\psi
+ g_+\bar{\psi}\Gamma^5P_+\psi
\right]\ \ \  \ \label{gpm}\ee
For $g_+=g_-$ and $f_+=f_-$, the deformation is Lorentz invariant and the boundary term reduces to \eqn{g1}. The advantage of parameterizing the deformation in this way is apparent when we insist that the resulting boundary condition, ${\cal M}\psi=0$, fixes only half the degrees of freedom where 
\be {\cal M} = \Gamma^4 + f_-P_-+f_+P_++g_-\Gamma^5P_-+g_+\Gamma^5P_+\nn\ee
The now familiar requirement that ${\rm rank}\,{\cal M}=2$ decouples the two sectors, leaving
\be f_+^2+g_+^2=1\ \ \ {\rm and}\ \ \ f_-^2+g_-^2=1\nn\ee
A similar decoupling occurs in the vacuum RG flow equation. The beta-function for $f_+$ is
\be \epsilon\frac{\partial f_+}{\partial\epsilon} = -2mL (1-f_+^2)\nn\ee
and the same equation holds for $f_-$. 
We learn that the fixed points are given by $f_+=\pm 1$ and, independently, $f_-=\pm 1$. The Lorentz invariant conformal field theories correspond to $f_+=f_-=+1$ and $f_+=f_-=-1$. The two other choices, $f_+=-f_-=\pm 1$, are the non-relativistic, scale invariant theories described in \cite{flatband}.

\para 
In the presence of sources, the non-relativistic RG flows are more complicated. As we saw for previous examples, the breaking of  translational invariance by the sources  induces higher derivative double trace operators along the flow. These are no longer constrained to be Lorentz invariant, meaning that, in addition to the two operators in \eqn{goingsouth}, we should also consider $h_\pm\bar{\Psi}\partial_t\gamma^0(1\pm i\gamma^0)\Psi$ and $p\bar{\Psi}\partial_i\gamma^i\Psi$, $i=1,2$, together with descendant operators that derive from acting with $\partial_t^2$ and $\partial_i\partial^i$. The resulting boundary terms  are
\be  S_B = \frac{i}{2}\int_{r=\epsilon} \sqrt{-\gamma}\left[f_\pm\bar{\psi}P_\pm\psi + g_\pm\bar{\psi}\Gamma^5P_\pm\psi + i\omega h_\pm\bar{\psi}\Gamma^0P_\pm \psi + ip \bar{\psi}k_i\Gamma^i \psi
 -\bar{\eta}\psi - \bar{\psi}\eta\right]\label{goodgod}\nn\ee
where we have introduced the notation of an implicit sum over $+$ and $-$ subscripts. 
Here all coefficients $f_\pm$, $g_\pm$, $h_\pm$ and $p$ are functions of $\omega^2$, $\vec{k}^2$ and $\epsilon$. The Neumann boundary condition is ${\cal M}\psi = \eta$, where ${\cal M}$ can be read off from the boundary term above. The complication arises when we impose the familiar condition that ${\rm rank}\,{\cal M}=2$. The spatial gradient terms, captured by the function $p(\omega^2,\vec{k}^2;\epsilon)$, couple the $P_+$ and $P_-$ sectors.  The result can be expressed by the requirement that ${\rm det}\,{\cal M}$ has two roots, where
\be {\rm det}\,{\cal M} &=& (f_-^2+g_-^2-\omega^2h_-^2-1)(f_+^2+g_+^2-\omega^2h_+^2-1) \nn\\ &&\ \ \ + 2p^2\vec{k}^2(f_+f_-+g_+g_--\omega^2h_+h_--1) + p^4\vec{k}^4\label{amen}\ee
This imposes two conditions which determine $f_\pm$ in terms of the other functions. In contrast to the relativistic deformations, this constraint is left implicit: we do not have explicit expressions for $f_+$ and $f_-$. 

\para
The condition ${\cal M}\psi = \eta$ can be manipulated to give the following two constraints on the spinors at $r=\epsilon$, 
\be \bar{\psi}(f_+P_++g_+\Gamma^5P_++i\omega h_+\Gamma^0P_+)\psi &=&\frac{1}{2}(\bar{\eta}P_+\psi + \bar{\psi}P_+\eta) - \frac{i}{2}p\bar{\psi}k_i\Gamma^i\psi\nn\\
\bar{\psi}(f_-P_-+g_-\Gamma^5P_-+i\omega h_-\Gamma^0P_-)\psi&=&\frac{1}{2}(\bar{\eta}P_-\psi + \bar{\psi}P_-\eta) - \frac{i}{2}p\bar{\psi}k_i\Gamma^i\psi \label{sodthis}\ee
We use these in evaluating the RG flow equation \eqn{rgflow} which is given by 
\be 
 && \epsilon\frac{\partial f_\pm}{\partial \epsilon}\bar{\psi}P_\pm\psi +  \epsilon\frac{\partial g_\pm}{\partial \epsilon}\bar{\psi}\Gamma^5P_\pm\psi +i\omega \epsilon\frac{\partial h_\pm}{\partial \epsilon}\bar{\psi}\Gamma^0P_\pm\psi + i \epsilon\frac{\partial p}{\partial \epsilon}\bar{\psi}k_i\Gamma^i\psi- \epsilon\frac{\partial \bar{\eta}}{\partial \epsilon}\psi - \epsilon\bar{\psi} \frac{\partial \eta}{\partial \epsilon} \nn\\ && \ \ \ = 2i\epsilon\bar{\psi}\kslash\psi - 2mL \bar{\psi}\psi -\frac{3}{2}(\bar{\eta}\psi + \bar{\psi}\eta)
\nn\ee 
Further invoking the constraints \eqn{sodthis} to eliminate two bilinears (most usefully $\bar{\psi}P_+\psi$ and $\bar{\psi}P_-\psi$) we find the beta function for the sources,
\be \epsilon P_+\frac{\partial\eta}{\partial\epsilon} = \left(\frac{3}{2}+\frac{mL}{f_+}+\frac{\epsilon}{2f_+}\frac{\partial f_+}{\partial\epsilon}\right)P_+\eta\nn\ee
together with the same expression with $+\rightarrow -$.  Similarly, the running of the operator coefficients are governed by
\be f_+\epsilon \frac{\partial g_+}{\partial \epsilon}-g_+\epsilon \frac{\partial f_+}{\partial \epsilon} &=& 2mLg_+
\nn\\ f_+\epsilon\frac{\partial h_+}{\partial \epsilon} - h_+\epsilon\frac{\partial f_+}{\partial \epsilon} &=& 2\epsilon f_++2mLh_+\nn\ee
both of which come with $+\rightarrow -$ counterparts. These equations  all take the same form as the relativistic beta-functions \eqn{sourcebeta}, \eqn{betamax} and \eqn{vhs}. The difference comes only in the  beta function for $p$, the coefficient of the spatial gradient term, which mixes the two sectors
\be f_+f_-\epsilon\frac{\partial p}{\partial \epsilon} - \frac{1}{2}f_-p\epsilon\frac{\partial f_+}{\partial \epsilon}-\frac{1}{2}f_+p\epsilon \frac{\partial f_-}{\partial \epsilon} = 2\epsilon f_+f_-+mLp(f_++f_-)\nn\ee
As in previous examples, these equations are to be understood with $f_+$ and $f_-$ fixed in terms of the other functions through the constraint \eqn{amen}.

\section{Spinors in AdS${}_5$}

We now turn to spinor fields in AdS$_{}5$.  There is no (interesting) analog of the Lorentz breaking deformation but we will describe versions of both the Dirac and Majorana deformations. However, the different gamma matrix structure means that each of these is different from its AdS${}_4$ counterpart. 

\para
We take the bulk gamma matrices to be
\be \Gamma^0 = \left(\begin{array}{cc} 0 & 1 \\ -1 & 0 \end{array}\right),\  \Gamma^1 = \left(\begin{array}{cc} 0 & \sigma^1 \\ \sigma^1 & 0 \end{array}\right)\  ,\  \Gamma^2 = \left(\begin{array}{cc} 0 & \sigma^2 \\ \sigma^2 & 0 \end{array}\right)\  ,\ 
\Gamma^3 = \left(\begin{array}{cc} 0 & \sigma^3 \\ \sigma^3 & 0 \end{array}\right) \ ,\ \Gamma^4 = \left(\begin{array}{cc} 1 & 0 \\ 0 & -1 \end{array}\right)\nn\ee
where $\Gamma^4$ again corresponds to the radial direction. The bulk charge conjugation matrix is ${\cal C}=\Gamma^2$. In what follows we consider $d=3+1$ dimensional Majorana and Dirac deformations

\subsubsection*{Majorana Deformation}

A bulk Dirac spinor $\psi$ in AdS${}_5$ is dual to a Weyl fermion operator in the $d=3+1$ dimensional boundary. With just a single Weyl spinor in hand, only the Majorana deformation is available to us. This takes the form
\be \Delta {\cal L}_{Majorana} = \zeta{\Psi}^T(i\sigma^2) \Psi + {\rm h.c.} \nn\ee
>From the bulk perspective there is correspondingly a  unique Majorana deformation of the boundary conditions consistent with the symmetries. In the absence of sources,  it is
\be S_B=  \frac{i}{2}\int_{r=\epsilon} \sqrt{-\gamma}\left[f\bar{\psi}\psi + \frac{1}{2}g\bar{\psi}^{(c)}\psi +\frac{1}{2}g\bar{\psi}\psi^{(c)}\right]\nn\ee
Note that this is actually the same deformation that we wrote down for the Majorana deformation in AdS${}_4$; the seeming lack of $\Gamma^5$ matrix is compensated by the need to define the charge conjugate spinor $\psi^{(c)}={\cal C}\psi$ differently in the $d=3+1$ and $d=4+1$ dimensional bulk. We can immediately conclude that the beta function for $g$ is again given by \eqn{splash}. 

\para
The same story holds in the presence of sources. The necessary deformations coincide with those of \eqn{fml}, with the $\Gamma^5$ operators absorbed into the definition of ${\cal C}$. The beta functions therefore agree with those of the Majorana deformation in AdS${}_4$ which, in turn, agree with those of the Dirac deformation in AdS${}_4$. There is a very minor difference in the beta function of the sources since the dimension of spacetime affects the dimension of the operator. We have
\be \epsilon\frac{\partial \eta}{\partial\epsilon} = \left(2+\frac{mL}{f}+\frac{\epsilon}{2f}\frac{\partial f}{\partial\epsilon}\right)\eta\label{higherpower}\ee
The beta functions for $g$ and $h$ are identical to equations \eqn{betamax} and \eqn{vhs}. 

\subsubsection*{Dirac Deformation}

To construct a Dirac operator $\Psi_\alpha$, $\alpha = 1,2,3,4$ in the boundary, we must 
employ two Dirac spinors in the bulk which we denote as $\psi$ 
and $\chi$, both of mass $m$. The first two components of $\Psi$ are donated by $\psi$; the last two by 
$\chi^{(c)}$. Note the charge conjugation on $\chi$ which ensures the correct helicity 
of the boundary spinors; if $\psi$ corresponds to a right-handed spinor on the boundary 
then $\chi^{(c)}$ corresponds to a left-handed spinor\footnote{One cannot impose a Majorana condition on 
a single 5d Dirac spinor, but given a pair of spinors one can impose a symplectic Majorana condition. 
This corresponds to a Majorana fermion operator on the boundary.}. Both bulk spinors are given boundary 
conditions corresponding to the alternative quantization\footnote{Equivalently, one could consider two spinors in the bulk with opposite masses, one with standard quantization the other with alternative quantization.} so that all components of $\Psi$ 
have dimension $\Delta_-$. 

\para
The Dirac deformation, 
\be  \Delta {\cal L}_{Dirac} = i\zeta\bar{\Psi}\Psi \nn\ee
breaks chiral symmetry of the boundary theory where, in our conventions, $``\gamma^5"\equiv \Gamma^4$. From the bulk perspective, this chiral symmetry rotates $\psi \rightarrow e^{i\alpha}\psi$ and $\chi^{(c)} \rightarrow e^{-i\alpha}\chi^{(c)}$. Or, in other words, both $\psi$ and $\chi$ have the same charge. This chiral symmetry is broken by the boundary term,
\be S_B=  \frac{i}{2}\int_{r=\epsilon} \sqrt{-\gamma}\left[f_\psi\bar{\psi}\psi + f_\chi \bar{\chi}\chi + g\bar{\psi}^{(c)}\chi +g \bar{\chi}\psi^{(c)}\right]\nn\ee
with the resulting boundary conditions
\be (f_\psi{\bf 1}+\Gamma^4)\psi = -g\chi^{(c)}\ \ \ ,\ \ \ (f_\chi{\bf 1}+\Gamma^4)\chi = -g\psi^{(c)}
\nn\ee
The Lorentz invariant choice $f_\psi=f_\chi=f$ is consistent with these boundary conditions providing we impose $f^2 + g^2 = 1$. Inserting this in the RG flow equation  yields the now familiar beta function \eqn{splash}. 
%
%

\para
The story in the presence of sources is also a familiar one. We'll again restrict to the choice $f_\psi=f_\chi$ and  upgrade the boundary term to 
\be S_B=  \frac{i}{2}\int_{r=\epsilon} \sqrt{-\gamma}\left[f(\bar{\psi}\psi +  \bar{\chi}\chi) + g(\bar{\psi}^{(c)}\chi + \bar{\chi}\psi^{(c)}) + ih(\bar{\psi}\kslash\psi +\bar{\chi} \kslash\chi)\right. \ \ \ \ \ \ \ \ \ \ \nn\\ \left.- \bar{\eta}_\psi\psi - \bar{\psi}\eta_\psi -\bar{\eta}_\chi\chi - \bar{\chi}\eta_\chi\right]\label{lastone}\ee
with $f$, $g$ and  $h$ all functions of $k^2$. 
This give rise to the boundary conditions 
\be (f+\Gamma^4 +ih\kslash)\chi(k) + g\psi^{(c)}(-k) &=& \eta_\chi(k) \nn\\
(f + \Gamma^4 -ih\kslash)\psi(-k) + g\chi^{(c)}(k) &=& \eta_\psi(-k)\nn\ee
where, as in Section \ref{majsec}, we have made the $k$ arguments explicit because the minus signs are important. The requirement that these boundary conditions fix only half the spinor components reduces to the usual constraint \eqn{f15}, together with a relationship between the bulk sources
\be g\eta_\psi(-k)+(f+\Gamma^4 -ih\kslash)\eta_\chi^{(c)}(k)=0\nn\ee
To determine the beta functions, we follow the usual procedure: the RG flow equation \eqn{rgflow} is evaluated on the boundary term \eqn{lastone} and are subsequently reduced using an identity which relates various fermion bilinears. This identity can be derived from the boundary conditions and a little algebra. It reads
\be f\left[\bar{\psi}\psi+\bar{\chi}\chi\right] + g\left[\bar{\chi}\psi^{(c)}+\bar{\psi}^{(c)}\chi\right] + ih\left[\bar{\psi}\kslash\psi+\bar{\chi}\kslash\chi\right]= \frac{1}{2}
\left[
\bar{\psi}\eta_\psi +\bar{\eta}_\psi\psi+\bar{\chi}\eta_\chi+\bar{\eta}_\chi\chi
\right] 
\nn\ee
%
%
With this in hand, it is a simple matter to read off the beta functions. The two sources $\eta_\psi$ and $\eta_\chi$ both independently obey \eqn{higherpower}. Meanwhile, the beta functions $f(k^2)$, $g(k^2)$ and $h(k^2)$ are given by the usual suspects: equations \eqn{betamax} and \eqn{vhs}, subject to the constrain \eqn{f15}. 

\para
It is worthy of comment that, despite very different spinor structures,  Majorana and Dirac deformations in both $d=2+1$ and $d=3+1$ give rise to the same beta functions. It strongly suggests that this result is due solely to large $N$.

\section*{Acknowledgement}
Our thanks to Nick Dorey for useful large N discussions. JNL is supported by the 
Funda\c{c}$\tilde{\rm a}$o para a Ci$\hat{\rm e}$ncia e Tecnologia (FCT-Portugal) through the grant SFRH/BD/ 36290/2007 and partially supported by CERN/FP/116377/2010. DT is supported by the Royal Society and ERC STG grant 279943.

\end{document}